\documentclass[prx,aps,onecolumn,showpacs,nofootinbib]{revtex4-2}

\usepackage[dvips]{graphicx} 
\usepackage{amsfonts,amscd,amsmath,amsthm}
\usepackage{enumerate}
\usepackage{epsfig}
\usepackage{subfig}
\usepackage{diagbox}
\usepackage{xcolor}
\usepackage[colorlinks = true]{hyperref}
\usepackage{physics}
\usepackage{epstopdf}
\usepackage{framed}
\usepackage{multirow}
\usepackage{color}
\usepackage{comment}
\usepackage[ruled,vlined]{algorithm2e}
\usepackage[most]{tcolorbox}

\graphicspath{{./figure/}}

\usepackage{tikz}
\usepackage{tikzpeople}
\usepackage{tikzlings}
\usetikzlibrary{tikzmark, calc, fit, positioning}
\usetikzlibrary{shapes,arrows.meta}
\usetikzlibrary{quantikz}

\newtheorem{observation}{Observation}

\newtcolorbox[auto counter]{mybox}[2][]{
	enhanced,
	breakable,
	colback=blue!5!white,
	colframe=blue!75!black,
	fonttitle=\bfseries,
	title=Box \thetcbcounter: #2,#1
}

\begin{document}
	
\title{Steering Alternative Realities through Local Quantum Memory Operations}

\author{Xiongfeng Ma}
\email{xma@tsinghua.edu.cn}
\affiliation{Center for Quantum Information, Institute for Interdisciplinary Information Sciences, Tsinghua University, Beijing 100084, China}

\begin{abstract}
	Quantum measurement resolves a superposition into a definite outcome by correlating it with an observer's memory—a reality register. While the global quantum state remains coherent, the observer's local reality becomes singular and definite. This work introduces reality steering, a protocol that allows an observer to probabilistically access a different reality already supported by the initial quantum state, without reversing decoherence on the environment. The mechanism relies on locally erasing the ``which-outcome'' information stored in the observer's brain. Here, ``local’’ means operations confined to the observer’s memory, excluding the environment, which may be cosmically large. Reality steering nevertheless faces intrinsic constraints: successful navigation requires coherent participation from the observer’s counterparts across the relevant branches, and any transition is operationally indistinguishable from non-transition. After arriving in a new reality, all memory records are perfectly consistent with that reality, leaving no internal evidence that a switch occurred. This makes conscious confirmation impossible within standard quantum mechanics. We show that nonlinear operations beyond the standard theory could, in principle, enable verifiable and deliberate navigation. Our results shift multi-reality exploration from philosophical speculation toward a concrete---though fundamentally constrained---quantum-informational framework.
\end{abstract}

\maketitle
\tableofcontents
\clearpage

\section{Introduction}
Quantum theory presents a profound paradox: the microscopic world is governed by coherent superposition, while our macroscopic experience is one of definite and unique outcomes. The standard model of projective measurement resolves this by describing a unitary interaction that entangles a quantum system with a memory-bearing apparatus (e.g., a device or a brain) and an environment \cite{Zurek2003Decoherence, Zurek2009Quantum}. From the global, external perspective of the purified wavefunction, this process is reversible and all possible outcomes continue to exist in a vast superposition. This is the core insight of Everett's relative-state formulation \cite{Everett1957Relative}, which interprets measurement as branching into coexisting outcome realities. From this perspective, no outcome is ever truly eliminated. However, our focus departs from the metaphysical picture of many worlds and instead treats these branches as information-theoretic alternatives that may, in principle, be navigated through local operations.

From the local perspective of an observer, Bob, whose perception is tied to the state of his brain or memory device, a single, definite outcome is experienced. This is not a ``splitting'' of particles or the creation of new universes composed of matter. Rather, it is the stabilization of one possibility into a consistent, information-theoretically isolated state—a \emph{reality}. The term ``multiverse,'' while inspiring, is misleading as it suggests a generation of distinct, parallel universes existing elsewhere in spacetime. In contrast, we advocate for the term multi-reality to emphasize that these are not separate physical worlds but different, mutually inaccessible states of the same universe. An observation does not create a new universe of atoms and electrons; it selects one thread of reality from the superposition of possible states.

This multi-reality framework highlights a fundamental operational asymmetry: the transition from a coherent superposition of possibilities to a singular reality (via measurement or decoherence) is effortless, while the reverse process—recovering coherence of a macroscopic system—is, for all practical purposes, cosmically impossible. The reason is straightforward: to restore coherence one would need to gather and recombine every fragment of ``which-outcome'' information that has leaked into the environment. Such information may be irretrievably dispersed, for instance, through photons emitted into interstellar space or microscopic perturbations that spread beyond control. Once this information escapes, it cannot realistically be recollected, rendering the reversal of decoherence infeasible on a cosmic scale. This asymmetry motivates a fundamental question: \emph{Given that global coherence reversal is forbidden, is it possible to steer realities through local manipulations of the memory?}

In this work, we propose a protocol for \emph{reality steering} based on local operations. The central idea is that Bob is not permanently confined to their current reality. If the specific record of the which-outcome result—stored in a \emph{reality register} within the Bob’s brain—can be selectively erased through a local quantum operation, then upon subsequent interaction with the world Bob may probabilistically experience an alternative reality that was part of the original quantum support. This is not time travel: the past remains immutable. Rather, it is a navigation of the future toward a different, yet self-consistent, branch of the multi-reality.

To explore the feasibility of this protocol, we can model the brain's memory encoding as a process analogous to quantum secret sharing \cite{Cleve1999How}, in which information is logically distributed across many neurons. This makes targeted erasure a highly non-trivial logical operation. As in quantum error correction, however, one does not need to manipulate all neurons related to the target information; it may suffice to swap the states of a few neurons into an ancillary system, thereby erasing the relevant record from the brain. Critically, successful steering requires a \emph{coordination constraint}: all Bobs in the targeted set of realities must undergo identical procedures, as inconsistent participation prevents proper disentanglement from the environment. Of course, this presupposes the ability to identify precisely which neurons carry the information, a challenge we leave to future advances in neuroscience.

This reality steering, however, reveals a more fundamental challenge: \emph{conscious reality steering}. Even if the technical hurdles of neuronal manipulation are overcome, the protocol's success from Bob's perspective remains unverifiable. After the procedure, whether Bob has genuinely navigated to a new branch or remained in his original one, his memory and experiences will be perfectly self-consistent with the reality he finds himself in. There is no internal witness or ``meta-memory" of the transition; both outcomes present an identical, coherent lived experience. This operational indistinguishability means that conscious confirmation of a reality switch is impossible within the standard quantum framework.

\section{Reality Clinic: Reality Steering}
The central premise of our framework is that an observer’s perceived reality is determined by the information encoded in their brain—the reality register. Through the act of observation or measurement, this register becomes correlated (often entangled) with a quantum system, thereby stabilizing one particular outcome from a superposition of alternatives. Measurement does not destroy superposition globally. Rather, it partitions the universal wavefunction into distinct branches, each carrying a consistent record of experience. The observer, by virtue of holding one such record, becomes confined to a single, definite version of reality.

In what follows, we introduce the operational notion of reality steering, a protocol designed to explore whether an observer can---by local manipulations of their own memory state---probabilistically access an alternative branch of reality already contained within the quantum superposition. The protocol does not attempt to reverse decoherence globally. Instead, it leverages the possibility of locally erasing or redistributing information within the observer’s memory so as to reestablish coherence between branches from that observer’s limited perspective. While a full reversal of the measurement process is cosmically impossible---due to the uncontrolled dispersion of information into the environment---we argue that a localized reset of the observer’s reality register remains, in principle, feasible within the scope of quantum mechanics.

To make the discussion concrete, we introduce three principal agents and a controlled environment we call the \emph{reality clinic}. The quantum system under observation, $C$, is Schrödinger’s cat, initially prepared in the superposition
\begin{equation}
	\begin{split}
		\ket{\psi}_C &= \tfrac{1}{\sqrt{2}}\bigl(\ket{\text{alive}} + \ket{\text{dead}}\bigr)_C \\
		&= \ket{+}_C .
	\end{split}
\end{equation}
The observer, Bob, serves as both participant and patient. Upon observing the cat, his brain registers one definite outcome and becomes entangled with the system, effectively collapsing his personal reality to one branch, either ``alive'' or ``dead.'' Eve represents the environment: she may also interact with the cat or Bob, thereby storing redundant copies of the which-outcome information. Eve’s degrees of freedom include photons, air molecules, sound vibrations, or any external record that retains traces of the measurement outcome. Importantly, these environmental records lie beyond Bob’s or the clinic’s direct control.

Then, we introduce Alice, who operates the reality clinic. The clinic is a controlled quantum laboratory isolated from environmental disturbances, where Alice performs local quantum operations on Bob’s brain designed to erase or overwrite the outcome-dependent record---the very memory that anchors Bob to his current reality. In essence, Alice’s surgical procedure targets the logical qubits that constitute Bob’s reality register. By coherently removing this information and resetting Bob’s brain to a neutral (uninformative) quantum state, she enables the possibility that Bob, upon re-engagement with the external world, will probabilistically find himself correlated with an alternative branch of the universal wavefunction.

It is crucial to emphasize that Alice’s actions are strictly local: she does not manipulate the cat’s state, nor does she interact with Eve’s environmental records. Her operations are confined to Bob’s physical substrate of memory—neurons, synaptic states, or equivalent quantum degrees of freedom that encode his observed outcome. From a physical perspective, the clinic can be thought of as a region of precise quantum control, capable of isolating and manipulating on Bob’s internal degrees of freedom without interference from the outside world. Theoretically, it functions analogously to a quantum laboratory implementing local erasure, state swapping, and controlled unitary operations on a living cognitive system.

\begin{figure}[htbp]
	\centering
	\begin{tikzpicture}[>=latex]
		\node[bob, minimum size=.5cm] (bob) at (0,0) {Bob};
		
		\node[
		draw=black, thick,
		fill=gray!10,
		rounded corners,
		minimum width=1cm,
		minimum height=0.75cm
		] (box) at ($(bob.east) + (.8, -0.6)$) {};  
		
		\draw[white, line width=3pt] (box.north west) -- (box.north east);
		
		\cat[shift={(box.south)},scale=.3,contour=black];
		
		\draw[->, dashed, gray] (bob.mouth) -- ($(box.center) + (50:0.5cm)$)
		node[midway, above, gray,sloped, font=\small] {observe};
		
		\node[bob, minimum size=.5cm] (bob1) at ($(box.east |- bob.east) + (2, 1)$) {};
		\cat[shift={($(bob1.east) + (0.4, -.4)$)},scale=.3,3D,body=orange!50];
		
		\node[bob, minimum size=.5cm] (bob2) at ($(box.east |- bob.east) + (2, -1.2)$) {};
		\cat[shift={($(bob2.east) + (0.7, -.1)$)},scale=.3,back,rotate=90];
		
		\draw[->, gray,thick] (box.north east) 
		to[out=15,in=180] 
		node[midway, above, gray,sloped, font=\small] {alive-cat} 
		(bob1.west);
		
		\draw[->, gray,thick] (box.north east) 
		to[out=-15,in=180] 
		node[midway, below, gray,sloped, font=\small] {dead-cat} 
		(bob2.west);
		
		\node[surgeon,sword, minimum size=.5cm] (alice) at ($($(bob1.east)!0.5!(bob2.east)$) + (2.7, 0)$) {Alice};
		\node[bob, minimum size=.5cm] (bob11) at ($(bob1.east) + (3, 0)$) {};
		\draw[->, gray,thick] ($(bob1.east) + (1, 0)$) to node[midway, below, gray,font=\small] {to clinic} (bob11.west);
		
		\node[bob, minimum size=.5cm] (bob21) at ($(bob2.east) + (3, 0)$) {};
		\draw[->, gray,thick] ($(bob2.east) + (1, 0)$) to node[midway, above, gray,font=\small] {to clinic} (bob21.west);
		
		\node[bob, minimum size=.5cm] (bob3) at ($(bob.east) + (9.5, 0)$) {};
		\node[left=.5cm of bob3.west,anchor=east,gray]{disentangling};
		\draw[->, gray,thick] (bob11.east) to[out=15,in=180] (bob3.west);		
		\draw[->, gray,thick] (bob21.east) to[out=-15,in=180] (bob3.west);		
		
		\node[bob,monitor, minimum size=.5cm] (bob4) at ($(bob3.east) + (1, 0)$) {};
		\draw[->, gray,thick] (bob3.east) to node[midway, below=.2cm, gray,font=\small] {re-observe} (bob4.west);
		
		\node[bob, minimum size=.5cm] (bob12) at ($(bob3.east) + (3.5, 1)$) {};
		\cat[shift={($(bob12.east) + (0.4, -.4)$)},scale=.3,3D,body=orange!50];
		
		\node[bob, minimum size=.5cm] (bob22) at ($(bob3.east) + (3.5, -1.2)$) {};
		\cat[shift={($(bob22.east) + (0.7, -.1)$)},scale=.3,back,rotate=90];
		
		\draw[->, gray,thick] (bob4.east) 
		to[out=15,in=180] 
		node[midway, above, gray,sloped, font=\small] {alive-cat} 
		(bob12.west);
		
		\draw[->, gray,thick] (bob4.east) 
		to[out=-15,in=180] 
		node[midway, below, gray,sloped, font=\small] {dead-cat} 
		(bob22.west);
	\end{tikzpicture}
	\caption{Reality steering protocol. Bob observes Schrödinger's cat, branching into alive-cat or dead-cat realities via GHZ-type entanglement among the cat ($C$), his brain ($B$), and the environment ($E$), given in Eq.~\eqref{eq:GHZ}. Bob visits the reality clinic, where Alice performs local memory erasure via SWAP operations. After disentanglement, Bob reobserves the world, potentially accessing an alternative reality—for example, transitioning from the dead-cat to the alive-cat branch.}
	\label{fig:RealitySteering}
\end{figure}
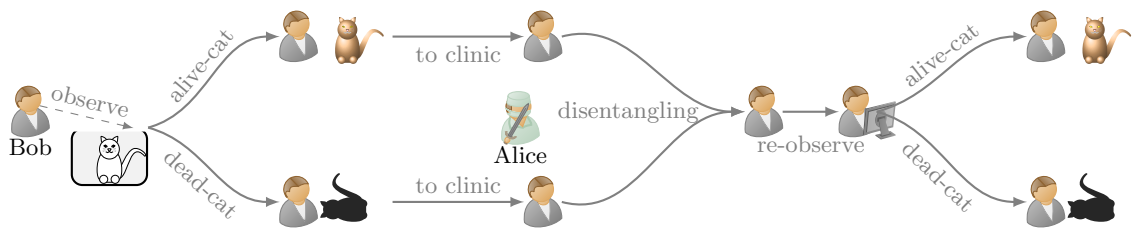

In this scenario, Bob, Alice, and Eve play conceptually distinct roles. Bob embodies the subjective perspective of reality—a local subsystem experiencing a definite world. Alice represents the agent of coherent intervention, probing the boundaries between classical observation and reversible quantum dynamics. Eve symbolizes the rest of the universe: the uncontrollable reservoir that irreversibly absorbs which-outcome information and thus enforces decoherence. Together, this triad defines the operational framework for analyzing the possibilities and limitations of steering between realities, as illustrated in Fig.~\ref{fig:RealitySteering}.

Let us first examine measurement, which can be understood as a process of information distribution. Rather than an abrupt quantum state collapse in the Copenhagen interpretation, measurement is described here as a unitary interaction $U_{\text{meas}}$ that distributes information about $C$ into other systems. Concretely, the cat becomes entangled with Bob’s memory $B$ and with Eve’s record $E$:
\begin{equation} \label{eq:GHZ}
	\begin{split}
		U_{\text{meas}} \ket{\psi}_C \ket{0}_B \ket{0}_E 
		&= U_{\text{meas}}\ket{+}_C \ket{0}_B \ket{0}_E \\
		&= \tfrac{1}{\sqrt{2}}\left(\ket{\text{alive}}_C \ket{a}_B \ket{a}_E 
		+ \ket{\text{dead}}_C \ket{d}_B \ket{d}_E\right),
	\end{split}
\end{equation}
where $\ket{a}_B, \ket{d}_B$ are orthogonal states of Bob's brain and $\ket{a}_E, \ket{d}_E$ orthogonal states of Eve. Operationally, this can be modeled as a sequence of CNOT gates: the cat acts as the control qubit, while Bob's brain and Eve's system are successive targets. The result is a GHZ-type entangled state \cite{greenberger2007goingbellstheorem} shared among $C$, $B$, and $E$.

Crucially, the CNOT operations are performed in a specific computational or natural basis determined by physical interactions. This basis is not arbitrary, but set by how the system couples to the apparatus and the environment—for instance through pointer-like degrees of freedom such as energy, photon number, or other robust collective variables. Such environmentally selected bases are stable under noise and enable rapid information dispersal, thereby driving decoherence. Their emergence is ultimately constrained by thermodynamics and the energetic cost of information processing \cite{Landauer1961Irreversibility}.

From the global perspective, superposition is preserved. Locally, however, any observer with access only to $C$ or $B$ perceives a mixed state, since tracing out $E$ removes off-diagonal coherence. This is decoherence: the which–outcome information has not vanished; it has been redundantly distributed across subsystems.

\begin{figure}[htbp!]
	\centering
	\subfloat[Bob and Eve observe independently]{
		\begin{quantikz}
			\lstick{$\ket{+}_C$} & \ctrl{1} & \ctrl{2} & \qw \\
			\lstick{$\ket{0}_B$} & \targ{} & \qw & \qw \\
			\lstick{$\ket{0}_E$} & \qw & \targ{} & \qw
		\end{quantikz}
	}
	\hspace{0.5cm}
	\subfloat[Bob observes, then informs Eve]{
		\begin{quantikz}
			\lstick{$\ket{+}_C$} & \ctrl{1} & \qw & \qw \\
			\lstick{$\ket{0}_B$} & \targ{} & \ctrl{1} & \qw \\
			\lstick{$\ket{0}_E$} & \qw & \targ{} & \qw
		\end{quantikz}
	}
	\hspace{0.5cm}
	\subfloat[Eve observes, then informs Bob]{
		\begin{quantikz}
			\lstick{$\ket{+}_C$} & \ctrl{2} & \qw & \qw \\
			\lstick{$\ket{0}_B$} & \qw & \targ{} & \qw \\
			\lstick{$\ket{0}_E$} & \targ{} & \ctrl{-1} & \qw
		\end{quantikz}
	}
	\caption{Three equivalent ways to establish the entangled state that defines Bob’s reality. All yield the same GHZ state, distributing which–outcome information across $C$, $B$, and $E$, as in Eq.~\eqref{eq:GHZ}.}
	\label{fig:ghz_circuit}
\end{figure}
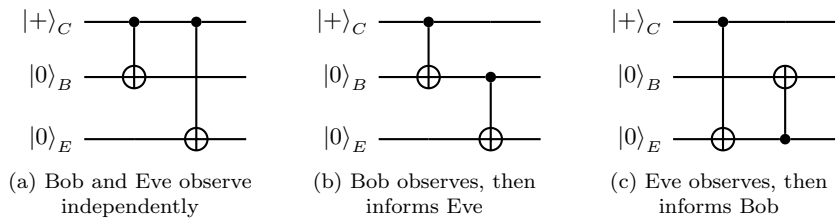

This entanglement can arise in several equivalent ways. Bob may look directly at the cat; Eve may observe the cat and then inform Bob; or both may observe independently. In all cases, the same GHZ correlation emerges, as illustrated in Fig.~\ref{fig:ghz_circuit}.

We can now state the \emph{cosmic impossibility of global reversal}. Suppose Bob observes and finds the cat alive. His reality has now collapsed into the alive-cat branch. Yet the which–outcome information is not confined to his memory $B$: Eve also possesses a correlated record, and the information has spread into countless environmental degrees of freedom---photons, air molecules, and sound waves. To truly reverse the measurement and restore the cat to its original superposition $\ket{+}_C$, one would need to apply the exact inverse unitary $U_{\text{meas}}^\dagger$ jointly to \emph{all three systems}, $C$, $B$, and $E$. Because $E$ is vast and uncontrollable, this would require recollecting every fragment of leaked information, including photons already traveling through deep space. If even a single copy is missing, reversal fails\footnote{This stringent requirement applies when information is copied in the computational basis. For more generic interactions, the threshold may be lower—for instance, for random unitaries, coherence recovery can succeed with access to roughly half of the system, as in quantum secret sharing \cite{Cleve1999How,li2025randomapproximatequantuminformation}. Even so, this remains cosmically large in realistic scenarios.}. Thus, the restoration of $\ket{+}_C$ is not merely impractical but \emph{cosmically impossible}.

This limitation motivates an alternative approach: rather than attempting cosmic-scale reversal, Alice can perform a localized intervention on Bob's reality register. When Bob enters the clinic, he leaves the cat and Eve's environmental records behind. Inside this controlled environment, Alice implements a quantum operation that erases the which-outcome information from Bob's brain state. After this reset, when Bob interacts with the world again, he may probabilistically find himself in a different branch of reality. The crucial insight is that, from Bob's subjective perspective, erasing his own memory record is sufficient; there is no need to modify the cat or the environment.

The core procedure of the reality clinic is to disentangle Bob's brain $B$ from the cat $C$ and environment $E$ by erasing the which-outcome record. This disentangling operation effectively reverses Bob's observation, as shown conceptually in Fig.~\ref{fig:ghz_circuit}. From a quantum information perspective, this is achieved by swapping the logical memory state into a fresh ancillary system $A$ (initialized to $\ket{0}$):
\begin{equation} \label{eq:SWAP}
	\begin{split}
		& \tfrac{1}{\sqrt{2}}\!\left(\ket{\text{alive}}_C \ket{a}_B \ket{a}_E
		+ \ket{\text{dead}}_C \ket{d}_B \ket{d}_E\right)\ket{0}_A \\
		& \xrightarrow{\ \text{SWAP}_{BA}\ }
		\tfrac{1}{\sqrt{2}}\!\left(\ket{\text{alive}}_C \ket{a}_A \ket{a}_E
		+ \ket{\text{dead}}_C \ket{d}_A \ket{d}_E\right)\ket{0}_B .
	\end{split}
\end{equation}
After this operation, $B$ is disentangled from $C$ and $E$, while the ancillary system $A$ now carries the memory record. From Bob's reset perspective, the cat (together with the rest of the world, $A+C+E$) returns to a superposition state.

\begin{figure}[htbp!]
	\centering
	\begin{quantikz}
		\lstick{$\ket{0}_A$} & \targ{} \gategroup[2,steps=2,style={inner
			sep=6pt,dashed}]{{\sc SWAP}} & \ctrl{1} & \qw & \rho_A \\
		\lstick{$\rho_B$} & \ctrl{-1} & \targ{} & \qw & \ket{0}_B
	\end{quantikz}
	\caption{Two–CNOT implementation of memory erasure. The first gate CNOT$_{BA}$ copies the record from Bob’s brain $B$ into the ancilla $A$. The second gate CNOT$_{AB}$ resets $B$ to $\ket{0}$ while preserving the record in $A$.}	
	\label{fig:SWAP}
\end{figure}
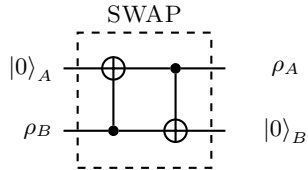

While a complete SWAP between $B$ and $A$ would normally require three CNOT gates, the initialization of $A$ to $\ket{0}$ allows for a more efficient ``move+reset'' protocol using only two CNOTs, as shown in Fig.~\ref{fig:SWAP}. The first CNOT$_{BA}$ copies the logical record from Bob's brain to the ancilla, while the second CNOT$_{AB}$ coherently resets $B$ to $\ket{0}$, thereby implementing the disentanglement operation of Eq.~\eqref{eq:SWAP}.

Physically, these two CNOT operations can be understood as follows. The first CNOT represents Alice's tool $A$ ``observing'' Bob's brain: $A$ becomes entangled with the $B+C+E$ system, effectively creating a coherent copy of Bob's memory. Informally, this corresponds to Bob telling Alice the outcome, but without losing his own memory. The second CNOT is substantially more demanding. Here, Alice must coherently flip Bob's memory conditional on the outcome being ``dead'', $\ket{1}$. This requires not only identifying the specific physical carriers of memory in Bob's brain (e.g., particular neurons or neural patterns) but also applying a controlled quantum operation using $A$ as the control. Thus, while the first CNOT resembles a relatively straightforward non-destructive readout, the second embodies the true challenge of coherent memory erasure.

The complete protocol for a \emph{reality clinic} can be summarized as follows:

\begin{mybox}[label={box:procedure}]{Reality Clinic Procedure}
	\emph{Step 0: Identify information carriers.} Locate the neurons or neural patterns encoding the cat’s outcome—analogous to identifying logical qubits in a distributed code.
	
	\begin{enumerate}
		\item 
		\emph{Initialization.} Prepare a fresh ancilla $A$ in $\ket{0}$ by a procedure independent of Bob’s observed outcome.
		
		\item 
		\emph{Coherent probing.} Couple $A$ to $B$ to coherently copy the record without destroying it (non-invasive quantum readout).
		
		\item  
		\emph{Memory reset.} Apply a controlled operation (with $A$ as control) that resets $B$ to $\ket{0}$ while preserving the record in $A$.
		
		\item 
		\emph{Reobservation.} After erasure, $B$ is disentangled from $A$, $C$, and $E$. Relative to $B$, the world is again in superposition; a new observation probabilistically places Bob into one of the branches (e.g., “alive”).
	\end{enumerate}
\end{mybox}

Mathematically, this protocol reduces to a pair of controlled operations, but physically it demands unprecedented precision. The clinic must isolate the specific neurons carrying the record, maintain coherence while coupling them to $A$, and then reset them without disturbing the rest of the brain. The fundamental limitations here are not cosmic in scale but rather stem from the extreme difficulty of local quantum control and fundamental restrictions such as superselection rules and conservation constraints. Since the memory information is typically distributed across many neurons---akin to quantum secret sharing \cite{Cleve1999How}---the required operation corresponds to a non-trivial logical transformation on the entire brain. The number of neurons the clinic must coherently control depends on the ``logical distance'' between the memory states $\ket{0}_B$ and $\ket{1}_B$. Insights from shallow quantum error correction \cite{liu2025approximatequantumerrorcorrection} suggest that significant, distributed control of neural degrees of freedom would likely be required for successful implementation.

\section{Feasibility and Theoretical Limits of Reality Steering}
While the reality steering protocol is mathematically coherent within standard quantum mechanics, its practical implementation faces several fundamental limitations rooted in the structure of the theory. First, quantum mechanics imposes strict coordination requirements: successful steering demands that all corresponding Bobs across the targeted branches participate and that Alice’s local operations be applied identically in each branch; otherwise, the memory–erasure step fails to disentangle Bob from the environment. Second, the linearity of quantum mechanics forbids both deterministic outcome control and any conscious verification of branch transitions: local operations cannot alter Born–rule probabilities, and even a perfectly executed steering operation leaves each Bob with internally consistent memories that provide no evidence that any transition has occurred. Both limitations are inherent to the linear theory. They could be circumvented only by introducing nonlinear evolutions capable of modifying branch probabilities—though such modifications would violate core physical principles, including no-signalling and the consistency of observer-independent reality.

\subsection{Coordination Constraints}
Before linearity imposes its restrictions on probability or verification, the protocol already encounters a more elementary obstacle: \emph{coordination across branches}. Even if perfect local erasure were physically implementable, it succeeds only when all participating branches are arranged in a branch-symmetric manner. Every Bob in the targeted family of realities must undergo the procedure, and Alice’s local operations must be applied identically in each reality. If either condition fails, the erasure operation does not remove the which-outcome information from Bob’s memory, leaving him entangled with the environment and preventing steering.

A direct demonstration of this constraint appears in the initialization of the ancillary system $A$. Alice must prepare $A$ in a way that is completely independent of Bob’s observed outcome. If Bob decides to visit the clinic only in certain branches—for example, only when he sees a dead cat—then $A$ becomes correlated with the branch structure, defeating the erasure. In such a flawed scenario, the joint state takes the form
\begin{equation} \label{eq:choose}
	\begin{split}
		& \tfrac{1}{\sqrt{2}}\!\left(\ket{\text{alive}}_C \ket{a}_B \ket{a}_E\ket{0,\text{off}}_A 
		+ \ket{\text{dead}}_C \ket{d}_B \ket{d}_E\ket{0,\text{on}}_A\right)  \\
		& \xrightarrow{\ \text{SWAP}\ }
		\tfrac{1}{\sqrt{2}}\!\left(\ket{\text{alive}}_C \ket{a}_B \ket{a}_E\ket{0,\text{off}}_A 
		+ \ket{\text{dead}}_C \ket{0}_B \ket{d}_E\ket{d,\text{on}}_A\right),
	\end{split}
\end{equation}
which leaves $B$ entangled with $C$ and $E$ and therefore blocks steering. Bob’s unilateral choice to seek the clinic in some branches but not others destroys the symmetry required for memory erasure.

This constraint extends to multi-outcome scenarios. Suppose the alive-cat and dead-cat realities comprise families of branches $\{\ket{a_i}\}$ and $\{\ket{d_j}\}$. If only the Bobs in the dead-cat branches participate, then
\begin{equation} \label{eq:choose_general}
	\begin{split}
		& \sum_i \ket{a_i}_C \ket{a_i}_B \ket{a_i}_E\ket{0,\text{off}}_A
		+ \sum_j \ket{d_j}_C \ket{d_j}_B \ket{d_j}_E\ket{0,\text{on}}_A  \\
		& \xrightarrow{\ \text{SWAP}\ }
		\sum_i \ket{a_i}_C \ket{a_i}_B \ket{a_i}_E\ket{0,\text{off}}_A
		+ \ket{0}_B \sum_j \ket{d_j}_C \ket{d_j}_E\ket{0,\text{on}}_A ,
	\end{split}
\end{equation}
with the result (up to normalization) that Bob can transition only among the participating dead-cat realities. The alive-cat branches remain inaccessible, demonstrating that coordination failure leads to irreversible restriction of navigable realities.

A second coordination requirement arises from the fact that different branches correspond to different physical states of the \emph{same} underlying particles. Alice must therefore implement \emph{exactly the same} quantum operation on Bob’s memory in every participating branch, even though the microscopic brain states may differ across those branches (for example, a Bob who is distressed or physiologically altered versus a Bob who is unharmed). At first sight, such uniformity seems unattainable: how can a single physical device act identically on memory states that are not microscopically identical?

The key point is that steering requires correct manipulation of the \emph{logical} information encoding the which-outcome record, not microscopic control of every neuron. If this information is distributed redundantly—analogous to quantum secret-sharing encodings—then Alice need only act on a sufficiently large subset of the relevant degrees of freedom to implement the logical SWAP that removes the record. In such encodings, many physical differences between branches are irrelevant to the logical operation, provided that Alice accesses a threshold subset, typically on the order of half of the encoded components\cite{li2025randomapproximatequantuminformation}.

These coordination constraints can be summarized succinctly:
\begin{observation}
	Reality steering requires that
	\begin{enumerate}
		\item all corresponding Bobs in the targeted branches participate in the steering operation, and
		\item Alice applies identical local quantum operations across those branches.
	\end{enumerate}
	Violation of either condition prevents Bob from becoming disentangled from the environment and therefore blocks access to the intended alternative realities.
\end{observation}

\subsection{Fundamental Limits: Probability Invariance and Unverifiability}
Even if the coordination requirements are satisfied, reality steering faces two fundamental limitations inherent to linear quantum mechanics. These limitations do not concern the ability to execute the protocol, but rather the impossibility of (i) modifying branch probabilities and (ii) verifying, from within any single reality, whether steering has occurred.

The first limitation follows immediately from the no–signalling theorem. Local operations cannot change outcome probabilities on a remotely entangled system. For a bipartite state $\ket{\psi}_{BC}=c_0\ket{00}+c_1\ket{11}$, any local completely positive trace-preserving map $\Lambda_B$ acting on Bob leaves the cat’s reduced state invariant:
\begin{equation}
	\begin{split}
		\rho_C' &= \tr_B[(I\otimes\Lambda_B)\rho_{BC}]
		= |c_0|^{2}\ketbra{0} + |c_1|^{2}\ketbra{1} \\
		&= \rho_C = \tr_B[\rho_{BC}].
	\end{split}
\end{equation}
Thus the Born-rule probabilities $|c_0|^2$ and $|c_1|^2$ remain fixed, regardless of how sophisticated Alice's local erasure operation may be. Within linear quantum mechanics, reality steering can only reshuffle Bob among branches with their original probabilities; deterministic outcome control is impossible.

A second limitation is the \emph{impossibility of conscious verification}. Even if steering succeeds globally—meaning the universal wavefunction has been coherently transformed—Bob’s subjective experience inside any single branch cannot reveal this fact. This is not a philosophical subtlety but a sharp quantum-information–theoretic constraint:
\begin{observation}
	No local measurement on Bob's degrees of freedom can distinguish between successful steering and no intervention.
\end{observation}

Formally, let $\rho_B$ be Bob's local state without steering and $\rho'_B$ his state after successful steering. Quantum linearity ensures $\rho'_B = \rho_B$, rendering the two situations operationally indistinguishable. Consequently, no internal experience or experimental procedure available to Bob can certify a reality transition. All memories, perceptions, and physical records remain perfectly self-consistent within his current reality. There exists no meta-memory, detectable disturbance, or empirical signature that could reveal the steering event. From Bob's perspective, a genuine steering event and a placebo are indistinguishable.

Together, these constraints imply that within linear quantum mechanics, reality steering affords neither probabilistic advantage nor empirical confirmation. The protocol is mathematically well-defined, but it lacks any operational witness accessible to an individual observer. Verifiable or controllable navigation of alternative realities would therefore require moving beyond the standard quantum theory framework.

These limitations naturally lead to a deeper question: are they fundamental necessities of quantum physics, or merely consequences of its linear form? If Bob were permitted to employ genuinely nonlinear operations at the level of quantum  evolution, the situation would change dramatically. Nonlinear transformations---which cannot be expressed as linear completely positive maps---can, in principle, rebalance the weights of different branches of a superposition, thereby overcoming both the invariance of Born-rule probabilities and the impossibility of verification. As shown in the Appendix \ref{app:nonlinear}, such nonlinear dynamics would allow deliberate branch selection and empirical confirmation: Bob could bias himself toward particular outcomes and, over repeated trials, observe statistical deviations from the Born rule, thereby obtaining conscious evidence that steering has occurred.

However, these expanded capabilities come at profound conceptual cost. Any probability-altering nonlinear operation necessarily violates the no-signalling principle and disrupts the internal consistency structure that standard quantum mechanics guarantees between observers. In this sense, linearity functions as a safeguard for branch autonomy, preventing any observer from exerting operational control over the larger superposition. While nonlinear extensions remain speculative, they illuminate the boundary between what quantum theory permits and what it forbids, revealing a deep connection between linearity, the stability of physical reality, and the limits of subjective experience.

\section{Discussion and Outlook}
We have introduced reality steering, an operational framework in which an observer can, in principle, navigate among quantum branches by locally erasing the which–outcome information encoded in their brain—the reality register. Such steering is achieved entirely through local operations that disentangle the observer from environmental records while preserving global unitarity. Although the past remains causally fixed, resetting the observer’s local state enables renewed interaction with the global superposition, allowing probabilistic access to alternative realities already present in the initial quantum state.

The core mechanisms of reality steering can be explored through controlled laboratory experiments. Mesoscopic quantum platforms—including superconducting qubits, trapped ions, and photonic networks—can generate GHZ-type entanglement and implement local erasure protocols. Quantum-eraser–style experiments may reveal coherence revival after targeted memory-reset operations, while appropriately designed temporal-correlation tests could witness the quantum character of the erasure process. These implementations would validate the theoretical framework without requiring biological consciousness.

Nonetheless, fundamental physical constraints severely limit what reality steering can achieve. The linearity of quantum theory, encoded in the no-signalling principle, prevents any local intervention from modifying Born-rule branch probabilities. Additional restrictions arise from physical considerations such as superselection rules associated with conserved quantities, the thermodynamic cost of erasure, and the structure of environmental decoherence. The effective computational bases that underlie perception may therefore be deeply shaped by thermodynamics and environmental interactions.

Extending these ideas to biological systems presents substantial challenges and opportunities. A central open problem is to develop physically realistic models of neural memory compatible with quantum control. This requires identifying the scales at which neural structures could maintain coherence, understanding how information is distributed across neuronal populations, and determining whether any neural degrees of freedom could plausibly carry entangled information relevant to perception. Equally important are questions about the resources required for coherence protection, the logical architecture of neural encoding, and the minimal control complexity needed for meaningful outcome modification. Advances on these fronts would clarify whether memory-level quantum operations—even in idealized or approximate forms—could interface with subjective experience or remain exclusively within the domain of engineered quantum systems.

\appendix
\section{Beyond Linear Quantum Mechanics} \label{app:nonlinear}
The only way to circumvent the limitations imposed by standard quantum mechanics---both the invariance of Born--rule probabilities and the impossibility of conscious verification---is to relax the linear structure of the theory itself. A reality clinic capable of deliberate or empirically verifiable branch selection would require a genuinely \emph{nonlinear} evolution acting on Bob’s subsystem, thereby stepping outside standard quantum mechanics.

Nonlinear modifications to the Schrödinger equation have been studied for decades and are known to generate qualitatively new phenomena, such as state-dependent evolution and, in many formulations, superluminal signalling \cite{GISIN19901}. Importantly, ``nonlinear'' does not automatically imply superluminal behaviour: for example, antilinear maps of the form, $U(c_0\ket{\psi} + c_1\ket{\phi}) = c_0^* U\ket{\psi} + c_1^*U\ket{\phi}$, are non-linear (as maps on state space) but preserve branch probabilities and therefore cannot enable controllable reality selection. In this work, we use \emph{linear} versus \emph{nonlinear} specifically to distinguish whether a local operation can modify branch weights; any such probability-altering transformation necessarily conflicts with the no-signalling principle.

To illustrate how nonlinearity would enable steering, consider again the GHZ-type state from Eq.~\eqref{eq:GHZ},
\begin{equation} \label{eq:GHZnonlinear}
	\ket{\Psi}_{CEB}
	=
	c_0 \ket{0}_C \ket{0}_E \ket{0}_B
	+
	c_1 \ket{1}_C \ket{1}_E \ket{1}_B ,
\end{equation}
where $\ket{0}$ denotes the alive-cat reality, $\ket{1}$ denotes the dead-cat reality, and $|c_0|^2 + |c_1|^2 = 1$.

Suppose Alice can implement a nonlinear, state-dependent operation acting only on Bob’s memory $B$. A simple example is a \emph{nonlinear filter} specified by a single (non-unitary) operator on $B$,
\begin{equation}
	K_\lambda = \ketbra{0} + \lambda \ketbra{1},
\end{equation}
where $\lambda>0$ sets the relative weighting of the two logical states. The associated transformation on the global state is
\begin{equation}
	\frac{(I_{CE}\otimes K_\lambda) \rho_{CEB} (I_{CE}\otimes K_\lambda)^\dagger}
	{\tr\left[(I_{CE}\otimes K_\lambda)\,\rho_{CEB}\,(I_{CE}\otimes K_\lambda)^\dagger\right]} .
\end{equation}
where the denominator introduces the essential nonlinearity: for $\lambda\neq 1$, this cannot be written as a linear completely positive trace-preserving map map.

Apply this operation to the pure entangled state in Eq.~\eqref{eq:GHZnonlinear}. The transformed and renormalized state is
\begin{equation}
	\frac{
		c_0 \ket{0}_C \ket{0}_E \ket{0}_B
		+
		\lambda c_1 \ket{1}_C \ket{1}_E \ket{1}_B
	}{
		\sqrt{|c_0|^2 + \lambda^2 |c_1|^2}
	}.
\end{equation}
Bob’s local nonlinear operation has therefore changed the cat’s outcome probabilities to
\begin{equation}
	\begin{split}
		P'_{cat}(0) &= \frac{|c_0|^2}{|c_0|^2 + \lambda^2 |c_1|^2}, \\[3pt]
		P'_{cat}(1) &= \frac{\lambda^2 |c_1|^2}{|c_0|^2 + \lambda^2 |c_1|^2}.
	\end{split}
\end{equation}
For $\lambda\neq 1$, these differ from the original Born probabilities $|c_0|^2$ and $|c_1|^2$. Bob’s \emph{local} nonlinear operation on $B$ has altered the statistics of the remote system $C$, explicitly violating the no-signalling theorem. In principle, Bob could bias himself toward the $\ket{0}$ branch using $0<\lambda<1$ or toward the $\ket{1}$ branch using $\lambda>1$. This would amount to tunable, active reality selection.

Such nonlinear dynamics would also resolve the verification problem inherent to the linear theory. Repeated application of such an operation would produce outcome frequencies that deviate systematically from Born–rule predictions, allowing a single observer to gather statistical evidence of branch bias. This feature sharply contrasts with the linear theory, where all observable data within any reality remain perfectly self-consistent and reveal no trace of a transition.

These nonlinear capabilities, however, come with a profound conceptual cost. Any probability-altering nonlinear operation inherently conflicts with the no-signalling principle and undermines the internal consistency guaranteed by standard quantum mechanics. Allowing such dynamics would reshape the operational and causal structure of the theory, erasing the autonomy of realities and permitting controllable inter-reality influence. For these reasons, the nonlinear mechanisms discussed here should be regarded as theoretical extensions used to delineate the boundary of what standard quantum mechanics forbids, rather than as physically established processes.

\section*{Acknowledgments}
This work is dedicated to my beloved sister, Liping. I deeply cherish her companionship. May she rest in peace in this reality, and live long and prosper in others. 

The author thanks L.~Huang and C.~Li for insightful discussions. 
This work was supported by the National Natural Science Foundation of China (Grants No. 12174216 and No. 12575023) and the Quantum Science and Technology-National Science and Technology Major Project (Grants No. 2021ZD0300804 and No. 2021ZD0300702).

\bibliographystyle{apsrev}
\bibliography{./tex/bibReality.bib}

\end{document}